\documentclass[showpacs,twocolumn,amssymb,amsmath,prb,aps]{revtex4}% Physical Review B
\usepackage[pdftex]{graphicx}

\newcommand{\etal}{\textit{ et. al.}}

\begin{document}
\title{Orientational Order in Liquids upon Condensation in Nanochannels: An Optical Birefringence Study on Rod- and Disc-like Molecules in Monolithic Mesoporous Silica}
\author{Matthias~Wolff}
\affiliation{Experimental Physics, Saarland University, D-66041 Saarbr\"ucken, Germany}
\author{Klaus~Knorr}
\affiliation{Experimental Physics, Saarland University, D-66041 Saarbr\"ucken, Germany}
\author{Patrick~Huber}
\affiliation{Experimental Physics, Saarland University, D-66041 Saarbr\"ucken, Germany}
\author{Andriy~V.~Kityk}
\affiliation{Institute for Computer Science, Czestochowa University of
Technology, Aleja Armii Krajowej 17, 42-200 Czestochowa, Poland}

\date{\today}

% General Remarks
% Our observations with respect to FB in agreement with MD simulations on experiments (Findenegg !!!). Orientational order in the first layers prevails in simulations. Remove all the details regarding the analysis. Focus on a discussion of the decay of the orientational disorder in the light of the literature and a possible influence on the adsorption/desorption hysteresis. First rod-like (comparison with hughe literature on calamatic liquid crystals), secondly disc-like molecules (comparison with MD simulations on benzene and fluorobenzene). Highlight the combination of monolithic membrane with high resolution birefringence in order to extract data on molecular orientational order

\begin{abstract}
We present high-resolution optical birefringence measurements upon sequential filling of an array of parallel-aligned nanochannels (14~nm mean diameter) with rod-like (acetonitrile) and disc-like (hexafluorobenzene) molecules. We will demonstrate that such birefringence isotherms, when performed simultaneously with optically isotropic and index-matched counterparts (neopentane and hexafluoromethane), allow one to characterize the orientational state of the confined liquids with a high accuracy as a function of pore filling. The pore condensates are almost bulk-like, optically isotropic liquids. For both anisotropic species we find, however, a weak orientational order (of a few percent at maximum) upon film-condensation in the monolithic mesoporous membrane. It occurs upon formation of the second and third adsorbed layer, only, and vanishes gradually upon onset of capillary condensation. Presumably, it originates in the breaking of the full rotational symmetry of the interaction potential at the cylindrical, free liquid-vapor interface in the film-condensed state rather than at the silica-liquid interface. This conclusion is corroborated by comparisons of our experimental results with molecular dynamics simulations reported in the literature.
\end{abstract}


\maketitle

\section{Introduction}
Liquids confined in tiny pores a few nanometers across can have properties which differ markedly from their bulk counterparts \cite{LitIntroExp, LitIntroTheo}. Both the interaction with the confining walls and the bare geometrical confinement can induce altered molecular packings, layering, changes in the molecular dynamics as well as preferred collective orientational order in the case of liquids consisting of anisotropic building blocks.

Whereas both altered packing and layering \cite{HarvardLMs, Volkmann2002} as well as orientational order \cite{Partay2005, Gramzow2007} has been established and systematically probed for a sizeable number of molecular liquids confined in a semi-infinite manner at the free surface of liquids and solids, such order phenomena upon mesopore confinement are still elusive. In fact, except for the well documented cases of strongly anisotropic mesogenes, which form nematic phases\cite{LiqCompGeometry}, only a few experimental evidences \cite{Korb1996, Valiullin2006} of such orientational order in mesopores have been reported so far for anisotropic building blocks, that form solely isotropic bulk liquids. 

Due to the complex pore structure, tortuous pore geometries and unknown pore wall/liquid interaction parameters it is particularly challenging to examine orientational order in mesopores. Nowadays, however, the advent of tailorable mesoporous substrates with aligned, non-interconnected tubular pores allows one to study this phenomenology. Here we present a high-resolution optical birefringence study upon sequential filling of an array of parallel-aligned silica nanochannels via the vapor phase. We will demonstrate that such birefringence isotherms, when employed simultaneously for optically isotropic and anisotropic building blocks (with identical refractive index), allow one to characterize the orientational state of the confined liquid with a high accuracy as a function of pore filling. 

\begin{figure*}[htbp]
\includegraphics*[width=1.8\columnwidth]{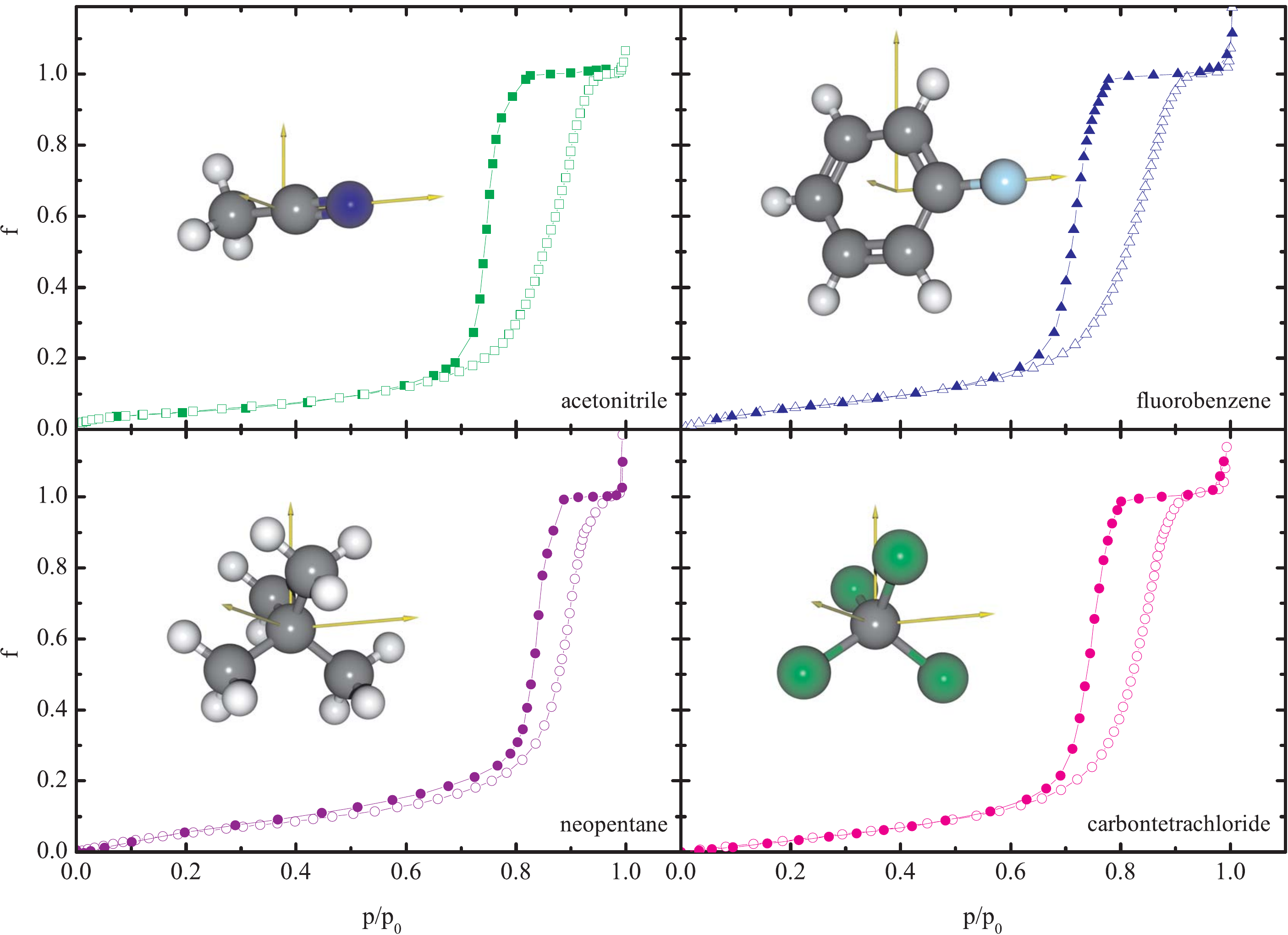} \caption{Volumetric sorption isotherms of the four molecular species invesitgated (as indicated in the figure), recorded at $T=$281~K in a monolithic mesoporous silica membrane. Open and solid symbols refer to ad- and desorption, respectively. The ray-tracing pictures illustrate the four molecular structures. The arrows represent the orientation of the principal axes of the polarizability tensors. Their lengths are proportional to the magnitude of the polarizabilities in those directions.} \label{fig:isotherms}
\end{figure*}

We have chosen the disc-like fluorobenzene (C$_6$H$_5$F, referred hereafter as FB) combined with its optically isotropic counterpart tetrachloromethane (CCl$_{4}$, TCM) and the rod-like acetonitrile (CH$_{3}$CN, AN) in combination with its optically isotropic partner neopentane (C$_{5}$H$_{12}$, NP), see Fig. \ref{fig:isotherms} for illustrations of the molecular structures. The bulk liquids of FB and TCM have practically identical refractive indices, and so do AN and NP (see Table 1).

The birefringence of these liquids have been measured as a function of fractional filling $f$ of the nanochannel array over the full range from the empty matrix ($f$=0) to the completely filled case ($f$=1). For each pair the experimental data on the anisotropic system are compared with the data of the isotropic system in order to extract a parameter typical of the collective orientational order of the molecules as a function of pore filling.

In a former study we employed an identical technique to obtain information on confined liquid n-hexane. However, we did not perform a comparison with an isotropic molecular counterpart, and hence had to use an effective medium model in order to scrutinize for any collective molecular orientation upon pore-condensation. It is understood that such a comparison can be affected by all kind of differences between the ideal model assumption and the real pore system, most prominently by a non-cylindrical pore shape and pore wall roughness. For n-hexane we did not find any clear cut evidence of a collective orientational order. This will be contrasted by the systems studied here.

\section{Experimental}
Mesoporous Si has been prepared by electrochemical etching of a highly $p$-doped (100) wafer with a resistivity of 0.01 $\Omega \cdot$cm by applying an electrical current of 12 mA/cm$^2$ for 8~hours. The resulting porous layer was released from the bulk Si underneath, by increasing the etching current by a factor of ten. Subsequently the porous Si sheet was thermally oxidized under standard atmosphere. The oven temperature was ramped upwards from 100$^o$C to 700$^o$C over 36~hours, kept at 700 $^o$C for 48 hours and finally ramped downwards to room temperature over 12 hours. After this procedure the membrane was entirely oxidized. Standard analysis of a $N_2$ sorption isotherm, measured at a temperature of 77~K, gives a porosity of 41\% and a pore diameter $2R_0$ of 14~nm. The sheet is 385 $\mu$m thick, hence the aspect ratio of the pores is $\sim$ almost 2.5 $\cdot$10$^5$. Electron micrographs of the channels in the original porous silicon membranes \cite{Gruener2008} indicate sizeable $1.0\pm0.5$~nm mean square deviations of their surfaces from an ideal cylindrical form. Presumably, the roughness in the oxidized samples are of similar magnitude. In order to minimize the influences of possible pore geometry variations on our experimental results upon usage of different samples, the identical sample was employed throughout all experiments.

Because of the perpendicular alignment of the channels' long axes to the surface, the wafer is an effective medium that shows an uniaxial
"shape" anisotropy, even in case the constituents have isotropic permittivities and refractive indices. Such examples of shape anisotropy have been observed in various systems, such as aligned carbon nanotube films \cite{Heer}, porous Si \cite{Gen,Kunz} or semiconductor nanowires \cite{Mus}. 

We have used a high-resolution optical polarization method for the accurate determination of the retardation $R$ at an incident angle of 33.7 deg. Details of the experimental setup can be found in Refs. \cite{Skarabot, Kityk2008, Kityk2009}. The samples have been kept in a closed optical chamber, held at a constant temperature ($T=$ 281 K) and small volumetrically controlled portions of gas have been subsequently added  (for adsorption) or removed (for desorption) from the chamber by means of a gas handling system equipped with a capacitive membrane gauge of 100~mbar full scale. The equilibration time was one hour for each step. The pore condensate is specified by the reduced pressure $p=P/P_{\rm 0}$ and filling fraction $f=N/N_{\rm 0}$. $P_{\rm 0}$ is the saturated vapor pressure of bulk hexane at given temperature $T$, $N$ is the number of adsorbed molecules in the porous substrate and $N_{\rm 0}$ is the number of molecules necessary for a complete filling of the pores. Optical birefringence measurements were only possible upon adsorption in the hysteretic part of the isotherm. Upon desorption a sizeable light scattering typical of a coarsening of the pore condensate upon desorption (see a discussion of this phenomenology in Ref. \cite{Kityk2009} and references therein) hampered polarimetry measurements. 
%\begin{sidewaystable}
\begin{table*}
\begin{center}
\begin{tabular}{c|c|c|c|c|c}
\hline\hline
\multicolumn{3}{c|}{}&\multicolumn{3}{c}{refractive indices}\\\hline
{\bf molecule}&
{\bf polarizability $ \alpha$}&
{\bf vapor pressure $ p_0$ ($8$)}&
$\bar{ n}$&
$ n_a$,$ n_b$,$ n_c$&
$ n_o$,$ n_e$\\\hline
\hline
CCl$_4$ -- tetrachlormethane -- TCM&
$\begin{pmatrix} 7.474 & 0.0 & 0.0 \\ 0,0 & 7.474 & 0,0 \\ 0.0 & 0.0 & 7.474 \end{pmatrix}$&
68.1 mbar&
1.459&
isotrope&
\\\hline
C$_6$H$_5$F -- fluorobenzene -- FB&
$\begin{pmatrix} 10.501 & 0.0 & 0.0 \\ 0.0 & 10.744 & 0.0 \\ 0.0 & 0.0 & 3.258 \end{pmatrix}$&
43.6 mbar&
1.463&
$\begin{matrix} n_a = 1.584\\ n_b = 1.601 \\ n_c = 1.161 \end{matrix}$&
$\begin{matrix} n_o = 1.593\\ n_e = 1.161 \end{matrix}$
\\\hline
\hline
C$_5$H$_{12}$ -- neopentane -- NP&
$\begin{pmatrix} 8.013 & 0.0 & 0.0 \\ 0.0 & 8.013 & 0.0 \\ 0.0 & 0.0 & 8.013 \end{pmatrix}$&
960.5 mbar&
1.342&
isotrope&
\\\hline
CH$_3$CN -- acetonitrile -- AN&
$\begin{pmatrix} 2.581 & 0.0 & 0.0 \\ 0.0 & 2.581 & 0.0 \\ 0.0 & 0.0 & 5.048 \end{pmatrix}$&
51.8 mbar&
1.346&
$\begin{matrix} n_a \sim n_b = 1.248 \\ n_c = 1.524 \end{matrix}$&
$\begin{matrix} n_o = 1.248\\ n_e = 1.524 \end{matrix}$\\\hline
\hline
\end{tabular}
\end{center}
\label{tab1}
\caption[]{Saturated vapor pressures (at $T=$281~K) and optical data sets of the molecules investigated \cite{PolTensor}.}
\end{table*} 
%\end{sidewaystable}
The observed retardation $R$ has been converted into the optical birefringence $\Delta n$ by means of the formalism described in detail in Refs. \cite{Kityk2008, Kityk2009}.

\section{Results and discussion}
\subsection{Sorption Isotherms}
The volumetric sorption isotherms of all four molecular species have a shape typical of condensation of van-der-Waals molecules in mesoporous substrates, see Fig. \ref{fig:isotherms}. The initial reversible part is due to adsorption of a film on the pore walls. The hysteretic part between adsorption and desorption is characteristic of capillary condensation and evaporation. The subsequent plateau indicates the completion of pore filling, whereas the final increase is due to the condensation of vapor on the machined rough metal walls of the sample cell and finally to bulk condensation. According to theoretical models \cite{Ev,Saam} for a cylindrical pore of uniform cross section the adsorption branch should be vertical in the regime of capillary condensation. The finite slope of both branches is typical of a broad distribution of pore diameters. Because of the rather gradual increase of the adsorption branch, the filling fraction $f_{\rm c}$ at which capillary condensation starts cannot be unambigiuously determined. Judging from our experience with similar van-der-Waals systems condensed in more regular cylindrical pores, e.g. SBA-15, we will just refer to an approximate value of 0.5 for $f_{\rm c}$ \cite{LitIntroExp}. 

\subsection{Birefringence isotherms}
In Fig. \ref{fig:OptBir} the birefringence $\Delta $n of all four molecular species is plotted as a function of filling fraction $f$ as recorded upon liquid adsorption. The general shape of these curves (i.e. the finite starting value of $\Delta $n, the maximum at $f$ about 0.3, and the subsequent decrease to very low values at $f$ approaching 1) is similar to that observed previously for hexane \cite{Kityk2009}. It is consistent with what is expected from a model describing aligned pores filled with an isotropic medium in an optically isotropic matrix. Nevertheless there are slight differences $\delta$ between the $\Delta n(f)$ data of the optically anisotropic molecules and their isotropic counterparts, $\delta $=$\Delta n_{\rm aniso}-\Delta n_{\rm iso}$. In Fig. \ref{fig:orderparameter} we show $\delta $ as function of $f$ for both the AN/NP and the FB/TCM pair. 

For the AN, the molecule with the prolate polarizibility tensor, $\Delta n_{\rm AN} > \Delta n_{\rm NP}$, that is $\delta$ positive, whereas for FB with the oblate tensor, $\Delta n_{\rm FB}< \Delta n_{\rm TCM}$, $\delta <0$.  This is evidence of an intrinsic orientational order of the AN and FB pore liquids, beyond the optical ``shape'' birefringence brought forth by the parallel alignment of the pores. In the present context birefringence is sensitive to $\cos^2{\theta}$ where $\theta $ is the angle between the pore axis and the direction of maximum (minimum) polarisibility $\alpha$ of the AN (FB) molecule (We ignore the small difference between the components $\alpha_a$ and  $\alpha_b$ of FB). Moreover, given the small and rigid character of the molecules employed along with the weakness of any of the interfacial potentials when compared to the covalent intra-molecular bondings, we feel encouraged to believe that no additional anisotropies within the single molecules are induced by the interfacial potentials, and thus hamper our analysis.

An idea of the degree of the preferred orientation is obtained from the comparison with a calculation of the birefringence $\Delta n$ for complete orientational order of these pore liquids, that is $\theta =0$ (see Ref. \cite{Kityk2009} for details of the model employed). In order to account for the observed signs of $\delta$, the AN molecules are all oriented with their axis parallel to the pore axis, whereas for FB it is the direction of smallest polarizibility that is parallel to the pore axis. For both cases the calculated birefringence changes strongly and monotonously as function of $f$, starting at $\Delta n=$0.018 at $f=0$ and attaining a value of 0.17 at $f=1$ for AN and -0.27 for FB. Calling the difference between these calculated $\Delta n$-values and the experimental ones $\delta_{\rm max}$, the quantity $q=\delta $/$\delta_{\rm max} =  1/2 \left < 3 \cos^2{\theta}-1 \right >$ serves as orientational order parameter. The results on $q(f)$ are included in Fig. \ref{fig:orderparameter}.

The values of $q$ are small, the maximum value of $q$ is of the order of one percent for both systems, and even these small values are only observed in a restricted $f$-range and by no means in the full range of $f$. Thus, ignoring a quite unlikely, peculiar angle of collective orientational order (discussed below), we can conclude that the AN and FB pore liquids are almost bulk-like, optically isotropic liquids.   

\subsection{Discussion}

The residual preferred orientation can in principle result from a breaking of the full rotational symmetry of the bulk liquid by the pore walls and/or the liquid-vapor interface. 

\begin{figure*}[htbp]
\includegraphics*[width=1.8\columnwidth]{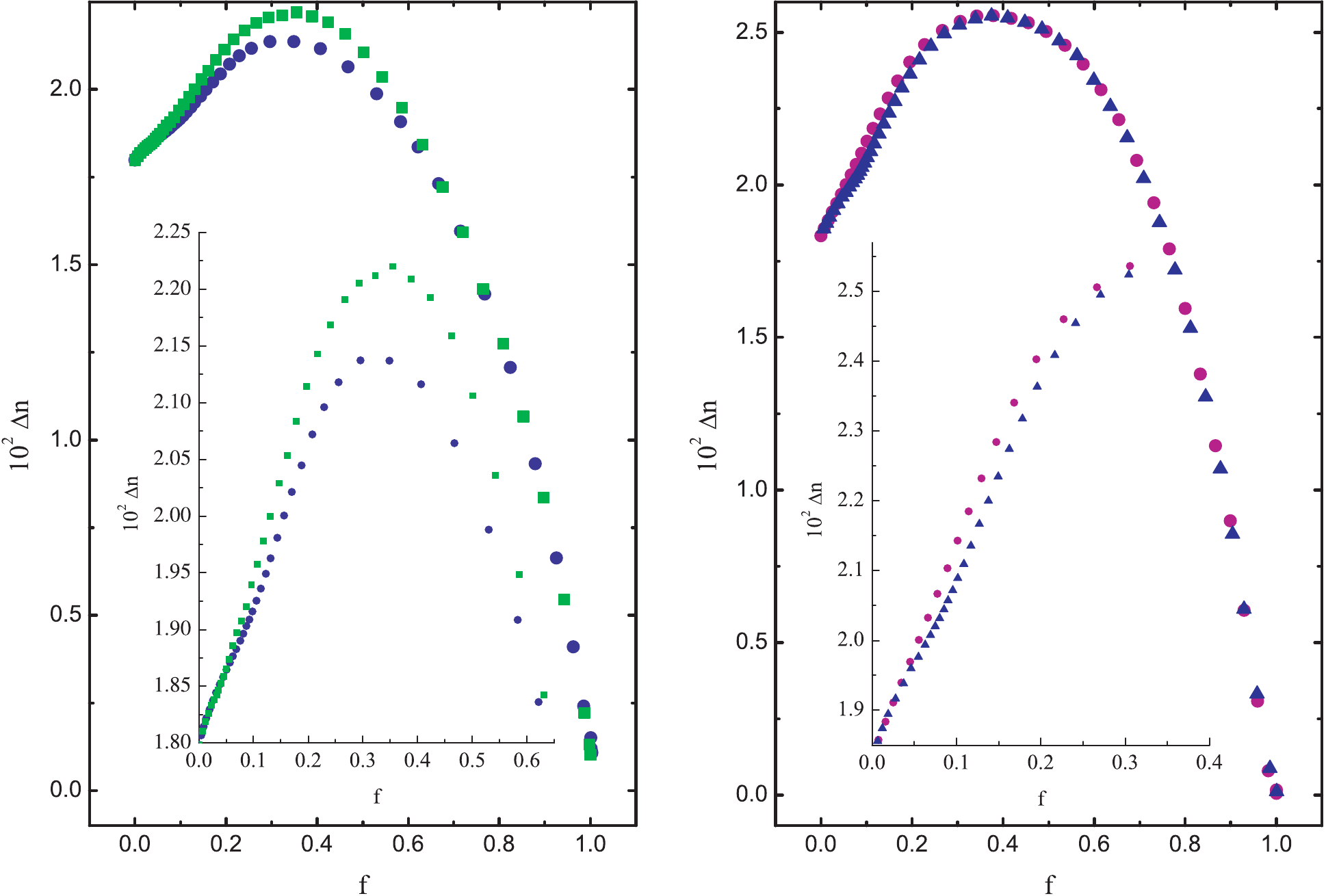}
 \caption{(Color online) The optical birefringence $\Delta n$ vs $f=N/N_0$ upon adsorption as determined from the measured phase retardation $R$ for NP (solid circles), AN (solid squares) in the left panel and for FB (solid triangles) and TCM (solid circles) in the right panel. The insets show the low-$f$ data on an expanded $\Delta n$-scale. } \label{fig:OptBir}
\end{figure*}

\begin{figure}[htbp]
\includegraphics[width=0.9\columnwidth]{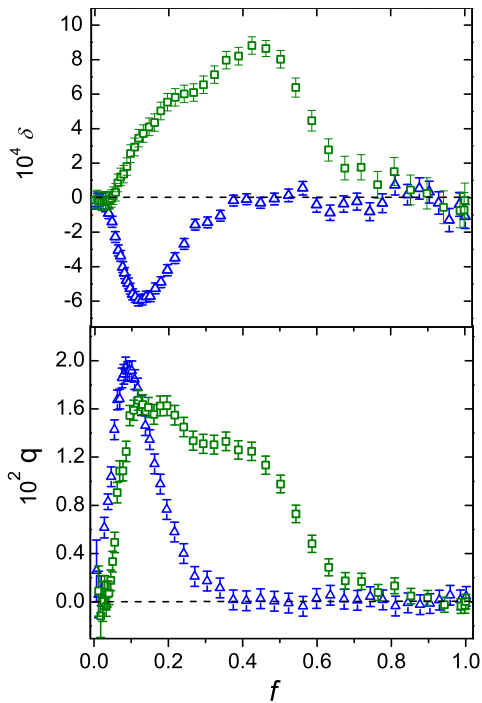}\caption{$\delta$ and $q$ vs. $f$ for the rod-like  acetonitrile (squares) and the disc-like fluorobenzene (triangles) condensed into a mesoporous silica membrane. \label{fig:orderparameter}} 
\end{figure}

We concentrate on AN for which the structure and the dynamics of the liquid at the liquid-vapor interface and at a silica wall have been studied by molecular dynamics simulations \cite{Hu2010, Morales2009, Paul2005}. According to this work the AN molecules of the liquid next to the wall show a broad distribution of tilt angles $\gamma $ with respect to the normal but with almost vanishing probability for $\gamma $=0, thus the molecules are preferentially aligned perpendicular rather than parallel to the wall. In the present experiment such preferred orientations should lead to a negative contribution to $\delta$ that should develop already in the submonolayer region of the sorption isotherm and persist up to complete filling. By contrast, we find vanishing values of $\delta$ in that regime. Hence, the experiment is compatible with an orientationally disordered state of the first molecules adsorbed. See the low starting values of $q(f)$, Fig. \ref{fig:orderparameter}, that arise from the fact that the birefringence curves $\Delta n(f)$ of AN and its isotropic reference liquid start with the same slope (Fig. \ref{fig:OptBir}). Of course pore walls of our samples are expected to be rough, both on a microscopic and on a mesoscopic scale. We anticipate that the resulting distribution of the surface normal can easily wipe out any preferred orientation with respect to the macroscopically defined direction of the pore axis. Note that, however, also for $\Theta=54.73$ deg ($3 \cos^2{\Theta} - 1=0$), the so-called magic angle, the resulting $q$ and thus also $\delta$ vanishes. Thus, in principle, we cannot exclude by our experiments values of $\gamma$ identical or close to the magic angle. Albeit, in the light of the aforementioned molecular dynamics studies this peculiar configuration is not expected \cite{Hu2010}.% We would like to stress, however, that our measurements in principal can not distinguish between those two cases.

In the following we argue that the finite, positive values of the orientational order parameter observed in the experiment are induced by the free surface of the pore liquid. According to the MD simulations the AN molecules at the liquid/vapor-interface have a slight preference for parallel alignment ($\gamma$=0). For cylindrical liquid/vapor-interfaces parallel to the pore axis this translates into positive values of $\delta$, in agreement with the experiment. One also expects that such interfaces are much smoother than the pore walls, thereby introducing less uncertainty when changing over from $\gamma$ to $\theta$. 

Finite values of $q$ occur for intermediate values of $f$ ranging from about $f=0.1$ to $f=0.6$. The delayed onset of $q(f)$ at low $f$ is related to the fact that the free surface of the liquid can only develop for liquid films on the walls that are sufficiently thick. According to Hu and Weeks \cite{Hu2010} the AN molecules within 2.5~nm from the silica wall are practically immobile. In the present samples with rough and highly curved pore walls the thickness, above which a free liquid/vapor surface can develop, is presumably even thicker. 
%\textbf{Also a vibrational sum frequency generation study of the free surface yields $\gamma = 40$ deg, a value which is smaller than, but close to, the magic angle. Such a preferred arrangement would still yield a positive, but a small $\delta$, and thus it may explain the weakness of the detected excess birefringence by a peculiar arrangement than by a weak general tendency for an orientational order}. 

The vanishing values of $q(f)$ beyond $f=0.7$ is a more problematic point. In the regime of capillary condensation ($f>f_c=0.5$) the area of the cylindrical liquid-vapor interfaces is expected to approach zero at $f=1$ in a linear fashion.  Thinking exclusively in such terms, $q(f)$ should show the same $f$-dependence. It is, however, well established that in the regime of capillary condensation vapor bubbles coexist with completely filled pore sections \cite{Naumov2008, Kityk2009}. Thus there are not only cylindrical but also meniscus-shaped liquid-vapor interfaces. With the molecules still preferentially parallel to the interface, the latter type interface supplies negative contributions to $\delta$. The fact that the partially filled silica pores with $f$-values within the capillary condensation regime are optically transparent suggests that the vapor bubbles are pretty small \cite{Naumov2008, Kityk2009}, but nevertheless it is surprising that the menisci are so abundant that their contribution to the preferred orientation of the molecules can compensate that of the cylindrical interfaces. Their great number could originate in the sizeable pore size variation and the corresponding plethora of nucleation sites for liquid bridges within the channels.

For FB finite values of $q$ are observed in an $f$-range considerably narrower than for AN (Fig. \ref{fig:orderparameter}). The negative values of $\delta$ are related to a preference of the normal of the aromatic ring pointing along the pore axis which translates into a perpendicular orientation with respect to the pore walls and to the cylindrical liquid/vapor interfaces. As for AN the vanishing value of $q(f)$ for f's, $f>0.3$, and for $f$ approaching 0, suggests that the present experiment does not produce evidence for pore wall induced preferential alignment, it rather suggests that any alignment observed is due to the liquid/vapor interface. Nevertheless it is remarkable that the intrinsic birefringence is confined to such a narrow $f$-range which corresponds to roughly the second and third monolayer adsorbed on the pore walls. 

In liquid C$_6$H$_6$ (and also in C$_6$F$_6$) the quadrupolar intermolecular interactions favor configurations where the symmetry axes of neighbouring molecules arrange perpendicular (\cite{Cabaco1997} and references therein). We assume that this interaction is also dominant in FB (in spite of its weak dipole moment) and that these configuration even prevail at the free liquid/vapor interface. In such a situation the combined contribution of neighbors to $q$ is just zero, thereby explaining why the FB molecules at the free surface do not contribute to the birefringence. We speculate that finite contributions can just occur in the special situation that such ''T''-like pair configurations have not yet developed. This is likely to be the case at $f$ about 0.1, where a fractional monolayer exists on the immobile molecules next to the wall.

Coasne et al. found in molecular dynamics simulations that the orientation of C$_6$H$_6$ at the silica wall is very sensitive to its hydroxilation \cite{Coasne2009}. Whereas for highly hydroxylated pore walls the ring is arranged perpendicular to the surface, partially hydroxylation leads to a disordered state in the first monolayer. Infrared spectroscopy measurements indicate that the pore walls are highly hydroxylated after the preparation procedure employed here \cite{Salonen1997, Kumar2010}. We should expect wall-induced orientational order, in contrast to our observation. Therefore, we have to resort to the assumption, similar to our considerations regarding the experiments on AN, that the orientational order in the wall proximity is washed out by roughness.

\section{Summary}
We presented a high-resolution optical birefringence study upon filling of a mesoporous monolithic silica membrane by four molecular species. The rod-like molecules acetonitrile and its optically isotropic and refrative-index-matched counterpartner neopentane as well as the disc-like fluorobenzene and its optically counterpartner tetrafluoromethane. A comparison of birefringence isotherms allowed us to gain detailed information on the collective order of the anisotropic building blocks, without having to resort to detailed model calculations on the changes of the optical birefringence upon capillary filling of the optically birefringent mesoporous matrix.

We find a very weak orientational order. The pore-condensed liquid is bulk-like, optically isotropic. Presumably, the weak orientational order is imposed by the breaking of the full rotational symmetry in the interaction potential at the cylindrical liquid/vapor interfaces upon film-condensation. Surpringly, upon onset of capillary condensation this preferred orientational order vanishes quickly which can only partially be understood by the vanishing liquid/vapor interfaces and the formation of concave menisci.

In general, the interpretation of our results significantly profited from molecular dynamics simulations in planar and cylindrical pore geometry, available in the literature. Thus our study also demonstrates that a thorough interpretation of  birefringence isotherms crucially relies on comparisons with such calculations. Alternatively comparisons with experimental results concerning molecular orientational order at the free, planar liquid surface, gained for example by vibrational sum frequency spectroscopy \cite{Partay2005}, will help to reduce the intrinsic ambiguities in the interpretation of the experimental method presented.

For the future, we suggest to systematically investigate the influence of pore wall roughness and chemistry, e.g. by silanization of the silica walls, or the usage of alternative, optically transparent mesoporous substrates, for example monolithic alumina membranes. The resulting variation of the pore wall/fluid interaction and reduced geometric pore wall irregularities may allow one to find and explore also collective orientational order at the fluid/solid interface, as expected by molecular dynamics studies.

\acknowledgments{This work has been supported by the German research foundation (DFG) within the project Hu850/3}.

\end{document}